\def\sgr{Sgr A$^{\ast}$}
\def\rfr#1{eq. (\ref{#1})}

\def\derp#1#2{\rp{\partial{#1}}{\partial{#2}}}
\def\dert#1#2{\frac{{{d}}{#1}}{{{d}}{#2}}}
\def\ros#1{{#1}}             

\def\virg#1{``#1''}

\def\eqi{\begin{equation}}
\def\eqf{\end{equation}}
\def\eqia{\begin{eqnarray}}
\def\eqfa{\end{eqnarray}}
\def\rp#1#2{{#1\over#2}} \def\lb#1{\label{#1}}

\def\bds#1{\boldsymbol{#1}}

\documentclass{aastex}
\usepackage{url}
\usepackage{amsmath,amsthm,amscd,amssymb}
\usepackage{latexsym}
\usepackage{graphicx,epsfig}

\RequirePackage{color}

\newcommand{\emaila}{lorenzo.iorio@libero.it}

\begin{document}

\title{Long-term classical and general relativistic effects on the radial velocities of the stars orbiting \sgr}
\shortauthors{L. Iorio}

\author{Lorenzo Iorio\altaffilmark{1} }
\affil{Ministero dell'Istruzione, dell'Universit\`{a} e della Ricerca (M.I.U.R.),\\
Fellow of the Royal Astronomical Society (F.R.A.S.).\\
 Viale Unit\`{a} di Italia 68, 70125, Bari (BA), Italy.}

\email{\emaila}

\begin{abstract}
We analytically work out the cumulative, i.e. averaged over one orbital revolution, time variations $\left\langle\dot v_{\rho}\right\rangle$ of the radial velocity $v_{\rho}$ of a typical S star orbiting the Supermassive ($M_{\bullet}\approx 10^6 M_{\odot}$) Black Hole (SBH) hosted by the Galactic Center (GC) in \sgr caused by several dynamical effects. They are the general relativistic gravitoelectromagnetic (GEM) fields of the SBH, its quadrupole mass moment $Q_2$, and a diffuse dark matter distribution around the SBH. All of them induce non-zero long-term radial accelerations proportional to the eccentricity $e$ of the orbit. By taking the S2 star, orbiting the SBH along a highly eccentric ($e=0.8831$) ellipse with a period $P_{\rm b}=15.9$ yr and semi-major axis $a=1031.69$ au, we numerically compute the magnitudes of its radial accelerations. The largest effects are due to the general relativistic Schwarzschild-like gravitoelectric (GE) field, with $\left\langle\dot v_{\rho}^{\rm (GE)}\right\rangle=8.2\times 10^{-5}\ {\rm m\ s}^{-2}$, and the diffuse material distribution, modeled with a Plummer-type mass density profile, with $\left\langle\dot v_{\rho}^{\rm (dm)}\right\rangle=3.8\times 10^{-6}\ {\rm m\ s}^{-2}$. The effects caused by the general relativistic Kerr-type gravitomagnetic (GM) field and by $Q_2$ are smaller by orders of magnitude. By assuming an uncertainty in measuring the radial velocities of about 15 km s$^{-1}$, the future accuracy in measuring $\left\langle\dot v_{\rho}\right\rangle$ can be evaluated to be of the order of $2.4\times 10^{-5}$ m s$^{-2}$ over an observational time span $\Delta t=20$ yr. Currently, the available radial velocity measurements cover just 7 yr.
\end{abstract}


\section{Introduction}
It is now widely accepted that the Galactic Center (GC) hosts a Supermassive Black Hole (SBH) \citep{Wol,Genzel,Ghez,nigri,Melia07}, whose position coincides with that of the radio-source Sagittarius A$^{\ast}$ (Sgr A$^{\ast}$) \citep{Reid07}.  The SBH has a mass of the order of $M_{\bullet}=4\times 10^6$M$_{\odot}$ \citep{Ghez,Gille,Gille2} and, consequently, a Schwarzschild radius  $r_g=0.084$ au. In its immediate vicinity a number of rapidly orbiting stars \citep{Pau} have been detected and tracked in the infrared since 1992 at the 8.2-m Very Large Telescope (VLT) on Cerro Paranal, Chile and the 3.58-m New Technology Telescope (NTT) on La Silla, Chile \citep{Ecka}, and since 1995 at the Keck 10-m telescope  on Mauna Kea, Hawaii \citep{Ghez98}. They are mostly main-sequence stars of spectral class B, and are naturally used as test particles for the gravitational potential in which they move. Particularly important is the bright star S2 (S0-2 in the Keck nomenclature), of spectral type B0-2 V, orbiting the SBH in $15.9$ yr along an  orbit with ellipticity $e=0.8831$ and semimajor axis $a=1031.69$ au \citep{Gille}. Indeed, a complete orbital revolution of it is now covered by the available data records  \citep{Gille,Gille2}.

If we look at the smallness of the ratio $\mathfrak{r}$ of the average distance $\left\langle r \right\rangle=a(1+e^2/2)$ to the Schwarzschild radius $r_g$ as an index of the importance of the Einstein's General Theory of Relativity (GTR) in several astronomical and astrophysical systems, it can be easily realized that the Galactic SBH and its stars is, in principle, an ideal local laboratory to put on the test GTR and other alternative theories of gravity. Indeed, by considering the Earth's artificial satellite LAGEOS \citep{LAGEOS}, the Sun's planet Mercury, the extrasolar planet WASP-19b \citep{WASP}, the double binary pulsar PSR J0737-3039A/B \citep{pulsar} and the  star S2 in \sgr, we have
\eqi
\begin{array}{lll}
\mathfrak{r}_{\rm LAGEOS}&=& 1.383319\times 10^9,\\ \\
\mathfrak{r}_{\rm Mercury}&=& 2.0023\times 10^7,\\ \\
 \mathfrak{r}_{\rm WASP-19b} &=& 8.74\times 10^5, \\ \\
 \mathfrak{r}_{\rm PSR\ J0737-3039A/B} &=& 1.15\times 10^5, \\ \\
 \mathfrak{r}_{\rm S2} &=& 1.7\times 10^4.
 \end{array}
 \eqf
 It can be noted that $\mathfrak{r}$ for S2 is one order of magnitude {smaller} than that for PSR J0737-3039A/B.
  In this respect, several authors \citep{Jaro,Fragile,Rubi,Weinb,Kran,Nuci,Will08,Preto,Khan,Merri} worked out with a variety of techniques and approximations the direct effects of GTR on different quantities characterizing the orbital motions about the SBH, mainly some Keplerian orbital elements. If, on the one hand, they are useful to give \virg{intuitive} valuable insights about the magnitude of the relativistic effects occurring in such a scenario, on the other hand they are not directly measurable.

Concerning the motion of the S stars around the SBH in the GC, the directly observable quantities are the astrometric measurements of their positions  in the sky in terms of right ascension $\alpha$ and declination $\delta$, and their radial velocities $v_{\rho}$.
Concerning the first kind of observations, according to \citet{Eise}, future astrometric measurements of S2 may bring its relativistic perinigricon\footnote{\citet{nigri} introduced in the scientific literature for the first time such a term for designing the pericenter $\omega$ in the case of a BH.  \ros{It comes from the Latin word \virg{$niger$}, meaning \virg{black}.} Before, the term \virg{perimelasma}  was used by the physicist G.A. \citet{Landis} in a science-fiction tale of him. \ros{Actually, the Greek word \virg{$\mu\acute{\varepsilon}\lambda\alpha\varsigma$} means \virg{black}, but,
on the other hand, \virg{melasma} ({\em Chloasma faciei}) is a medical term denoting a skin affection which consists of a tan or dark skin discoloration An earlier and somewhat more adequate term is \virg{peribothron}, coming from the Greek word \virg{$\beta\acute{o}\theta\rho o\varsigma$} for \virg{pit}. It is due to a suggestion by W.R. Stoeger to \citet{bothron}. After all, BHs may not be entirely \virg{black} since, under certain circumstances, they may radiate, so that a term recalling their deep gravitational well appears more suitable to characterize them.  Let us note that the definition of BHs
as \virg{frozen stars} by the former Soviet scholars
may have lead to the word \virg{peripekton}, from the Greek word \virg{$\pi\eta\kappa\tau\acute{o}\varsigma$} for \virg{frozen}. } } precession $\left\langle\dot\omega_{\bullet}\right\rangle$ into the measurability domain.
 Indeed, the perinigricon rate would be indirectly inferred from the corresponding apparent position shift. To this aim, it must also be considered that
such a shift is not as easily
detected as it may seem since it needs to be measured from
the same data from which also the orbital elements have to
be determined \citep{precizi}.
Thus, here we will focus on the dynamical effects directly caused by GTR and other competing classical forces on the radial velocity $v_{\rho}$. On the one hand, it will be possible to straightforwardly work out in an analytical way the net time variations of it averaged over one orbital revolution. This allows for a more direct and unambiguous confrontation of the theoretical predictions with the observations.  On the other hand, from a practical point of view the radial velocity data are easier to handle with respect to the astrometric observations. Indeed,  the inclusion of new data into pre-existent records needs no special care because the radial velocities refer to the Local Standard of Rest (LSR) \citep{Reid}. Instead, for the astrometric data it turns out that only an approximate realization of a \ros{common relative} reference frame is possible. It implies that the exact definition of the coordinates is a matter of each data analysis in such a way that simply merging two different sets of astrometric positions would yield incorrect results \citep{Gille2}.

In our calculation, we will proceed as follows.
For the sake of generality, let us assume that
an explicit, analytical expression is available for a given observable $Y$ in such a way that it
is function of all or some Keplerian orbital elements, i.e. $Y=Y(f,\{\kappa\})$, where $f$ is the true anomaly, and $\kappa$ denotes the ensemble of the Keplerian orbital elements explicitly entering $Y$ apart from the mean anomaly $\mathcal{M}$. Then, we straightforwardly compute its secular variation as the sum of two parts. The first one is purely Keplerian, and it vanishes over one orbital period $P_{\rm b}$. The second one is due to the non-Keplerian variations of all the  orbital elements induced by the dynamical perturbation considered. The total result is, thus,
\eqi \left\langle\dert{Y}{t}\right\rangle=\rp{\left\langle\Delta Y\right\rangle}{P_{\rm b}}=\left(\rp{1}{P_{\rm b}}\right)\int_0^{2\pi}\left[\derp{Y}{f}\dert{f}{\mathcal{M}}\dert{\mathcal{M}}{t}+\sum_{\kappa}\derp{Y}{\kappa}\dert{\kappa}{t}\right]\left(\dert{t}{f}\right) df.\lb{ypa}\eqf
In it,  ${d\mathcal{M}}/{dt}$ and $d\kappa/dt$ are the instantaneous variations\footnote{Actually, $d\mathcal{M}/dt$ is the sum of the Keplerian mean motion $n$ and a non-Keplerian term, as we will see later. Its Keplerian part  yields from \rfr{ypa} the Keplerian variation of $Y$.} of the Keplerian orbital elements computed with, e.g., the Gauss variation equations and evaluated onto the unperturbed Keplerian ellipse, while ${df}/{d\mathcal{M}}$ and $dt/df$ are the usual Keplerian expressions for such derivatives: see \rfr{yta} and \rfr{yga} below.

The paper is organized as follows. In Section \ref{accelez} we deal with some kinds of both classical and relativistic perturbing accelerations. The long-term effects caused by them on the radial velocity are analytically worked out in Section \ref{velazza}.
In Section \ref{numerizi} we perform numerical calculations by using the S2 star, and confront them with the present-day measurement accuracy. Section \ref{conclu} is devoted to the summarizing our findings.
\section{The perturbing accelerations}\lb{accelez}
\subsection{General treatment and overview}
Here we deal with a generic perturbing acceleration $\bds A$ induced by a given dynamical effect which can be considered as small with respect to the main Newtonian monopole $A_{\rm Newton}=GM_{\bullet}/r^2$, where $G$ is the Newtonian constant of gravitation and $r$ is the relative star-SBH distance. The stars orbiting the SBH are assumed test particles: their masses are about $m_{\star}\lesssim 10^{-5} M_{\bullet}$, and relativistic corrections to their internal structures are assumed to be too small to yield noticeable effects on their orbital motions.

First, $\bds A$ has to be projected onto the radial, transverse and normal orthogonal unit vectors $\bds{\hat{R}},\bds{\hat{T}},\bds{\hat{N}}$ of the co-moving frame of the test particle orbiting the central body acting as source of the gravitational field.
Their components, in cartesian coordinates of a reference frame centered in the primary, are  \citep{Monte}
\eqi \bds{\hat{R}} =\left(
       \begin{array}{c}
          \cos\Omega\cos u\ -\cos \lambda\sin\Omega\sin u\\
          \sin\Omega\cos u + \cos \lambda\cos\Omega\sin u\\
         \sin \lambda\sin u \\
       \end{array}
     \right)\lb{ierre}
\eqf
 \eqi \bds{\hat{T}} =\left(
       \begin{array}{c}
         -\sin u\cos\Omega-\cos \lambda\sin\Omega\cos u \\
         -\sin\Omega\sin u+\cos \lambda\cos\Omega\cos u \\
         \sin \lambda\cos u \\
       \end{array}
     \right)\lb{itrav}
\eqf
\eqi \bds{\hat{N}} =\left(
       \begin{array}{c}
          \sin \lambda\sin\Omega \\
         -\sin \lambda\cos\Omega \\
         \cos \lambda\\
       \end{array}
     \right)\lb{inorm}.
\eqf
In \rfr{ierre}-\rfr{inorm}, $\Omega,\omega,\lambda$ are the longitude of the ascending node\footnote{It is an angle in the reference $\{xy\}$ plane from the reference $x$ direction to the line of the nodes. }, the argument of pericenter, reckoned from the line of the nodes\footnote{It is the intersection of the orbital plane with the reference plane $\{xy\}$.}, and the inclination of the orbital plane to the reference $\{xy\}$ plane, respectively. In general, we will choose the unit vector $\bds{\hat{\rho}}$ of the line-of-sight, pointing from the object to the observer, to be  directed along the positive $z$ axis, so that the $\{xy\}$ plane coincides with the usual plane of the sky. With such a choice, corresponding to the frame actually used in data reduction \citep{Ghez}, $\lambda$ is the familiar $i$, and $\Omega$ is an angle in the plane of the sky counted from the mean vernal point at J2000 epoch along which the reference $x$ axis is customarily chosen; it is such a node which is actually determined from the observations \citep{Ghez,Gille,Gille2}. In other cases, in order to compute more easily certain dynamical perturbations, we will orient our frame with the $z$ axis aligned with the central body's proper angular momentum $\bds L$, so that the reference $\{xy\}$ plane will coincide with the body's equatorial plane. In this case, $\Omega$ will be an angle lying in such a plane, and it is not the one released in literature \citep{Ghez,Gille,Gille2}. Moreover, $u\doteq f+\omega$ is the argument of latitude.
Subsequently, the projected components of $\bds A$ have to be evaluated onto the Keplerian ellipse
\eqi r=\rp{p}{1+e\cos f},\ p\doteq a(1-e^2),\lb{rkep}\eqf where $p$ is the semilatus rectum and $a,e$ are the semi-major axis and the eccentricity, respectively.
The cartesian coordinates of the Keplerian motion in space are
 \citep{Monte}
 \begin{equation}
{\begin{array}{lll}
 x &=& r\left(\cos\Omega\cos u\ -\cos \lambda\sin\Omega\sin u\right),\\  \\
 y &=& r\left(\sin\Omega\cos u + \cos \lambda\cos\Omega\sin u\right),\\  \\
 z &=& r\sin \lambda\sin u,
\end{array}}\lb{xyz}
 \end{equation}
 while the cartesian components of the velocity can be obtained as
 \eqi
 \begin{array}{lll}
 v_x &=& \derp{x}{f}\dert{f}{t},\\ \\
 v_y &=& \derp{y}{f}\dert{f}{t},\\ \\
 v_z &=& \derp{z}{f}\dert{f}{t},\\ \\
 \end{array}
 \eqf
  in which $df/dt$ is given by \rfr{yga}.

Then,  $A_R,A_T,A_N$ are to be plunged into the right-hand-sides of the  Gauss equations for the variations of the Keplerian orbital elements. They are \citep{Roy,Soffel}
\begin{equation}
{\begin{array}{lll}
\dert{a}{t}&=& \rp{2}{n\sqrt{1-e^2}}\left[A_R e \sin f+ A_T\left(\rp{p}{r}\right)\right],\\ \\
\dert{e}{t}&=& \rp{\sqrt{1-e^2}}{na}\left\{ A_R\sin f +A_T\left[\cos f+\rp{1}{e}\left(1-\rp{r}{a}\right)\right]\right\},\\ \\
\dert{\lambda}{t}&=&\rp{1}{na\sqrt{1-e^2}}A_N\left(\rp{r}{a}\right)\cos u,\\\\
\dert{\Omega}{t}&=& \rp{1}{na\sqrt{1-e^2}\sin \lambda}A_N\left(\rp{r}{a}\right)\sin u,\\\\
\dert{\omega}{t}&=& -\cos \lambda\dert{\Omega}{t}+\rp{\sqrt{1-e^2}}{nae}\left[-A_R\cos f+A_T\left(1+\rp{r}{p}\right)\sin f\right], \\ \\
\dert{\mathcal{M}}{t}&=& n-\rp{2}{na}A_R\left(\rp{r}{a}\right)-\rp{(1-e^2)}{nae}\left[-A_R\cos f+A_T\left(1+\rp{r}{p}\right)\sin f\right],
\end{array}}\lb{Gauss}
 \end{equation}
 where  $n\doteq \sqrt{GM/a^3}$ is the Keplerian mean motion  related to the orbital period by $n=2\pi/P_{\rm b}$.

 As explained in the Introduction, the right-hand-sides of \rfr{Gauss}, computed for the perturbing accelerations of the dynamical effect considered, have to be inserted into the analytic expression of the time variation $dY/dt$ of the observable $Y$  of interest which, then, must be averaged over one orbital revolution according to \rfr{ypa} by means of \citep{Roy}
\eqi df =\left(\rp{a}{r}\right)^2\sqrt{1-e^2}d\mathcal{M},\lb{yta}\eqf
and
 \eqi dt = \rp{(1-e^2)^{3/2}}{n(1+e\cos f)^2}df.\lb{yga}\eqf
\subsection{The effect of general relativity}
In its slow-motion and weak-field approximation, GTR predicts that a slowly rotating central body of mass $M$ and proper angular momentum $\bds L$ induces two kinds of small perturbations on the otherwise Keplerian orbital motion of a test particle. The largest one is dubbed gravitoelectric (GE) \citep{Mash}, and depends only on the mass $M$ of the body which acts as source of the gravitational field. It is responsible of the well-known anomalous secular precession of the perihelion of Mercury of $43.98$ arcsec cty$^{-1}$ in the field of the Sun. There is also a smaller perturbation, known as gravitomagnetic (GM) \citep{Mash}, which depends on $\bds L$: it causes the Lense-Thirring \citep{LT} precessions of the node and pericenter of a test particle.
In the linearized gravitoelectromagnetic (GEM) approximation, the general relativistic perturbing acceleration $\bds A_{\rm GTR}$ to be added to the Newtonian monopole $\bds A_{\rm Newton}$ is \citep{Soffel}
\eqi \bds A_{\rm GTR}=-\bds E_g -2\left(\rp{\bds v}{c}\right)\bds\times\bds B_g,\lb{gemeq}\eqf
with
\eqi
\begin{array}{lll}
\bds E_g &=& -\rp{GM}{c^2 r^3}\left[\left(\rp{4GM}{r}-v^2\right)\bds r+4\left(\bds r\bds\cdot\bds v\right)\bds v\right], \\ \\
\bds B_g &=& -\rp{G}{cr^3}\left[\bds L - 3\left(\bds L\bds\cdot\bds{\hat{r}} \right)\bds{\hat{r}}\right],\lb{egos}
\end{array}
\eqf
where $c$ denotes the speed of light in vacuum.
In \rfr{gemeq}-\rfr{egos} $\bds E_g$ is the GE field, while  $\bds B_g$ is the GM one.
In regard to $\bds B_g$, in the case of a rotating BH, the existence of the horizon in the \citet{Kerr} metric, which describes the spacetime outside it, implies a maximum value for its angular momentum \citep{Bar1,Mel1}
\eqi L_{\bullet}^{(\rm max)}=\rp{M^2_{\bullet} G}{c}.\eqf If such a limit is actually reached or not by  astrophysical BHs depends on their accretion history \citep{Barde}. In fact, recent measurements of the spin of the  SBH in \sgr obtained in the context of discseismology by
means of newly detected quasi-periodic oscillations (QPOs) of radio emission point towards \citep{Kato}
\eqi \chi_{\bullet}\doteq\rp{L_{\bullet} c}{M_{\bullet}^2 G}=0.44\pm 0.08.\lb{spinbh}\eqf
\ros{
 \citet{Genznat} showed that for \sgr the spin parameter is $0.52\pm (0.1,0.08,0.08)$
or larger, if a QPO observed from \sgr in 2003 is of
 dynamical origin. On the other hand, it must noted that X-ray QPOs oscillations are rather disputed.
}
Actually, in the case of the S stars orbiting the SBH in \sgr, while the weak-field approximation is acceptable since for S2
 \eqi \rp{\left\langle U\right\rangle}{c^2}=\rp{GM_{\bullet}}{c^2}\left\langle\rp{1}{r}\right\rangle=\rp{GM_{\bullet}}{a c^2}=4\times 10^{-5}, \eqf
  the slow-motion approximation may be, in principle, less adequate. Indeed, the speed of, say, S2 at perinigricon is as large as {$2.6\%$} of the speed of light, while at aponigricon it is {$0.1\%$} of $c$. Thus, in the dynamical equation of motion of \rfr{gemeq} higher order relativistic corrections should be, in principle, taken into account \citep{Masho,Capozzo}. Anyway, as we will see, they induce negligible consequences on the dynamical effects we are interested in, given the present-day level of accuracy in S stars spectroscopy.

\subsubsection{The gravitoelectric, Schwarzschild-like perturbation}
By defining
\eqi \mathcal{R}_g\doteq \rp{GM}{c^2},\eqf the $R-T-N$ components of the  general relativistic GE perturbing acceleration $\bds E_g$ are \citep{Soffel}
 \begin{equation}
{\begin{array}{lll}
A_R^{(\rm GE)}&= & \rp{n^2 \mathcal{R}_g}{(1-e^2)^3}(1+e\cos f)^2(3+2e\cos f -e^2 +4e^2\sin^2 f), \\ \\
A_T^{(\rm GE)}&= & \rp{n^2 \mathcal{R}_g}{(1-e^2)^3}(1+e\cos f)^2 4e\sin f(1+e\cos f),\\ \\
A_N^{(\rm GE)}&=& 0.
\end{array}}\lb{ge}
 \end{equation}
 Note that $[\mathcal{R}_g]=$ L, so that $[n^2 \mathcal{R}_g]=$ L T$^{-2}$. Moreover, \rfr{ge} does not depend on the inclination of the orbit to the plane of the sky.
\subsubsection{The gravitomagnetic, Lense-Thirring-like perturbation}
The $R-T-N$ components of the Lorentz-like general relativistic GM perturbing acceleration  induced by the rotation of the central body with proper angular momentum $L$ are \citep{Soffel}
 \begin{equation}
{\begin{array}{lll}
A_R^{(\rm GM)}&= & \eta_g \cos\Psi(1+e\cos f), \\ \\
A_T^{(\rm GM)}&= & -\eta_g e \cos\Psi\sin f,\\ \\
A_N^{(\rm GM)}&=& \eta_g\sin\Psi(1+e\cos f)\left[2\sin u + e\left(\rp{\sin f\cos u}{1+e\cos f}\right)\right],
\end{array}}\lb{gm}
 \end{equation}
 with
 \eqi\eta_g\doteq\rp{\xi_g n}{a^2(1-e^2)^{7/2}}(1+e\cos f)^3,\eqf and \eqi \xi_g\doteq \rp{2 G L}{c^2}.\eqf
 Note that $[\xi_g]=$ L$^3$ T$^{-1}$, so that $[\eta_g]=$ L T$^{-2}$.
 In \rfr{gm} $\Psi$ is the angle between the orbital plane and the equatorial plane of the central body, i.e. $\bds L$ has been assumed directed along the positive $z$ axis so that the reference $\{xy\}$ plane coincides with the equatorial plane of the central body ($\lambda\rightarrow \Psi$).
 For equatorial orbits, i.e. for $\Psi=0$, $A_N^{\rm (GM)}=0$ and $A_R^{\rm (GM)}\neq 0, A_T^{\rm (GM)}\neq 0$. Instead, for polar orbits, i.e. for $\Psi=90$ deg, only the normal component does not vanish.
\subsection{The quadrupole mass moment}
The external gravitational field of a rotating body undergoes departures from spherical symmetry because of the distortion of its shape due to the resulting centrifugal force.
An oblate body of equatorial radius $R_e$ and adimensional quadrupole mass moment $J_2$ affects the orbital motion of a test particle with a  non-central perturbing acceleration \citep{Cunn,Vrbik}
\eqi \bds{A}^{(J_2)}=-\rp{3J_2 R_e^2 GM}{2r^4}\left\{\left[1-5(\bds{\hat{r}}\bds\cdot\bds{\hat{L}})^2\right]\bds{\hat{r}}+2(\bds{\hat{r}}\bds\cdot\bds{\hat{L}})\bds{\hat{L}}\right\},\lb{sgorbia}\eqf
where $\hat{\bds{L}}$ is the unit vector of the body's angular momentum, directed here along the positive $z$ axis so that the equatorial plane is the reference $\{xy\}$ plane (again, $\lambda\rightarrow\Psi$).
According to \rfr{ierre}-\rfr{inorm} and \rfr{xyz}, the $R-T-N$ components of \rfr{sgorbia}
 are
 \begin{equation}
{\begin{array}{lll}
A_R^{(J_2)}&\doteq & \bds{A}^{(J_2)}\bds\cdot\bds{\hat{R}} = -\rp{3 n^2  R_e^2 J_2}{8a(1-e^2)^4}\left(1+e\cos f\right)^4\left(1+3\cos 2\Psi +6\sin^2\Psi \cos 2u\right), \\ \\
A_T^{(J_2)}&\doteq &  \bds{A}^{(J_2)}\bds\cdot\bds{\hat{T}} = -\rp{3 n^2  R_e^2 J_2}{2a(1-e^2)^4}\left(1+e\cos f\right)^4\sin^2\Psi\sin 2u,\\ \\
A_N^{(J_2)}&\doteq &  \bds{A}^{(J_2)}\bds\cdot\bds{\hat{N}} = -\rp{3 n^2  R_e^2 J_2}{2a(1-e^2)^4}\left(1+e\cos f\right)^4\sin 2\Psi\sin u.
\end{array}}\lb{j2}
 \end{equation}
 Note that $[n^2 R^2_e a^{-1}]=$ L T$^{-2}$. For $\Psi=0$, i.e. for equatorial orbits, only the radial component is not zero. For polar orbits, i.e. for $\Psi=90$ deg, the normal component vanishes, contrary to the radial and transverse ones.

 Also a rotating BH should have a quadrupole mass moment, so that it affects the orbital motion of a distant test particle with a perturbing acceleration analogous to that of \rfr{sgorbia}. It is customarily to introduce a dimensional quadrupole parameter $Q_2$, $[Q_2]={\rm L}^5\ {\rm T}^{-2}$, in such a way that \citep{Will08} \eqi J_2 R^2_e GM\rightarrow Q_2\eqf throughout \rfr{sgorbia} and \rfr{j2}.
 According to the \virg{no-hair} or uniqueness theorems of GTR \citep{hair1,hair2}, an electrically neutral BH is completely characterized by its mass $M_{\bullet}$ and angular
momentum $L_{\bullet}$ only. As a consequence, all the multipole moments
of its external spacetime are functions of $M_{\bullet}$ and $L_{\bullet}$. In particular,
the quadrupole mass moment is
\eqi Q^{\bullet}_2=-\rp{L_{\bullet}^2 G}{c^2 M_{\bullet}}=-\chi^2_{\bullet} \rp{G^3 M_{\bullet}^3}{c^4}.\eqf
Thus, in the case of the SBH in GC \rfr{spinbh} yields
\eqi \left|Q^{\bullet}_2\right|=3.585\times 10^{45}\ {\rm m^5\ s^{-2}}. \lb{sbozo}\eqf
In terms of the adimensional coefficient $J_2$, by assuming the Schwarzschild radius $r_g$ for the equatorial radius $R_e$ of the BH, \rfr{sbozo} would correspond to
\eqi \left| J_2^{\bullet}\right|=4\times 10^{-2}.\eqf Just for the sake of a comparison, for the Sun, Jupiter and the Earth we have $J_2^{\odot}=2\times 10^{-7}$ from helioseismology \citep{sunj2}, $J_2^{\rm (Jup)}=1.46\times 10^{-2}$ from the flybys of some spacecraft \citep{jup230}, $J_2^{\oplus}=1.08\times 10^{-3}$ from the dedicated GRACE spacecraft \citep{grace}, respectively.
\subsection{Inner diffuse mass distribution}
It is well recognized that, in addition to the dynamical effects considered so far directly related to the SBH, also the impact of a diffuse cluster of non-luminous ordinary matter\footnote{\ros{In principle, one should also take into account the issue of non-baryonic dark matter cusps in relation to the merging history of galaxies \citep{cusp}. It is beyond the scopes of this work.}} around the BH due to massive remnants of various kinds \citep{Mor} should be taken into account. Indeed, reasoning in terms of the perinigricon, such an extended material component would induce a retrograde precession $\dot\omega_{\bullet}$ which may overwhelm the general relativistic GE one for certain values of its mass \citep{Rubi}.
Let us, now, work out in detail its dynamical effects.

Following \citet{Rubi}, \citet{Mou}, \citet{Gille2}, we adopt a Plummer density profile
\eqi\varrho_{\rm dm}(r)=\rp{3\mu M_{\bullet}}{4\pi d^3_{\rm c}}\left(1+\rp{r^2}{d_{\rm c}^2}\right)^{-5/2},\eqf
where the core radius is \citep{Gille2} \eqi d_{\rm c}= 15\ {\rm mpc},\lb{densita}\eqf
in agreement with the observed light profile \citep{Mou}, while $\mu$, the mass parameter, is the ratio of the total extended mass $M_{\overline{r}}$ at a given distance $\overline{r}$ to the central point mass.
\citet{Gille2}, with a fit involving S2 able to probe the mass enclosed between its aponigricon and  perinigricon, yield
\eqi\mu\leq 0.04.\eqf
\ros{
It should be mentioned that \citet{Mou}, with the position of \sgr as a free discrete input parameter, provided for the first time an upper limit of
\eqi \mu\leq 0.05.\eqf
}
From the Poisson equation
\eqi \bds\nabla^2 \mathcal{U}=\rp{1}{r^2}\rp{\partial}{\partial r}\left(r^2\derp{\mathcal{U}}{r}\right)=4\pi G\varrho\eqf for the gravitational potential $\mathcal{U}$, written in spherical coordinates since $\mathcal{U}=\mathcal{U}(r)$ in view of the fact that $\varrho=\varrho(r)$, the perturbing acceleration
\eqi \bds A_{\rm (dm)}=-\bds\nabla \mathcal{U}_{\rm  (dm)}=-\derp{\mathcal{U}_{\rm (dm)}}{r}\bds{\hat{r}}\eqf  easily follows. It turns out
\eqi
\begin{array}{lll}
A_R^{\rm (dm)} &=& -\rp{GM_{\bullet}\mu}{d^3_{\rm c}}\left(1+\rp{r^2}{d^2_{\rm c}}\right)^{-3/2}r, \\ \\
A_T^{\rm (dm)} &=& 0, \\ \\
A_N^{\rm (dm)} &=& 0.
\end{array}\lb{dmacc}
\eqf
\section{The radial velocity}\lb{velazza}
The basic observable in spectroscopic studies of binary systems  is the radial velocity $v_{\rho}$, i.e. the component of the velocity vector $\bds v$ of one of (or both) the system's partners along the line-of-sight whose unit vector $\bds{\hat{ \rho}}$ has been assumed directed along the $z$ axis.
\ros{Concerning the S stars orbiting the Galactic SBH, let us mention the possibility that some of them may actually be binaries themselves. In this case, additional \virg{noise} to the radial velocity data would be introduced. }
Its expression for unperturbed, Keplerian elliptic orbits, up to the velocity of the system's center of mass $v_0$,
can be obtained by using the
$z$ components of \rfr{ierre} and \rfr{itrav} with $\lambda\rightarrow i$,  and recalling that, for a Keplerian orbit \citep{Roy},
\eqi \bds v=v_R\bds{\hat{R}}+v_T\bds{\hat{T}}=\rp{na}{\sqrt{1-e^2}}\left[e\sin f\bds{\hat{R}}+(1+e\cos f)\bds{\hat{T}}\right].\eqf The result is
\eqi v_{\rho}=K\left[e\cos\omega+\cos(f+\omega)\right],\lb{radvel}\eqf
where $2K$ is the total observed range of radial velocity
defined by
\eqi K\doteq\rp{na\sin i}{\sqrt{1-e^2}}.\lb{kvel}\eqf
Note that \rfr{radvel} and \rfr{kvel} agree with the expressions given by, e.g., \citet{Ency} and \citet{Pad}.

Perturbing dynamical effects affect the radial velocity as well by inducing, in principle, a non-vanishing  net radial acceleration over one orbital period.
It can straightforwardly be worked out from \rfr{ypa} with $Y\rightarrow v_{\rho}$ by noting that, in this case,  the perturbations of all the six Keplerian orbital elements are involved. In this respect, a special care is required for the parameter $i$ entering \rfr{kvel}.
It is the angle between the unit vector $\bds{\hat{\ell}}$ of the orbital angular momentum and the unit vector $\bds{\hat{\rho}}$ of the line-of-sight pointing towards the observer. From
the spherical law of cosines
\citep{Gel,Zwi}
\eqi\cos B=\sin C\sin A\cos b-\cos C\cos A\lb{trigga}\eqf  with the identifications\footnote{In such a way,  $\Omega$ results to be prograde with respect to the orbital motion, i.e. $\Omega$ follows it, coherently with the definition of the longitude of ascending node. Moreover, it lies in the equatorial plane.} $A\rightarrow \Psi,B\rightarrow \pi-i,C\rightarrow I,b\rightarrow\pi-\Omega$,
 it turns out
\eqi\cos i=\sin\Psi\sin I\cos\Omega+\cos\Psi\cos I,\lb{spherical}\eqf
where $\Psi$ is the angle between $\bds{\hat{\ell}}$ and the unit vector $\bds{\hat{L}}$ of the central body's proper angular momentum, $I$ is the angle between $\bds{\hat{L}}$  and $\bds{\hat{\rho}}$, and $\Omega$ is the longitude of the ascending node defined from
\eqi\sin\Psi\sin I\cos\Omega
=(\bds{\hat{L}} \bds\times \bds{\hat{\ell}})
\bds\cdot
(\bds{\hat{L}} \bds\times \bds{\hat{\rho}}).\lb{megas}\eqf
From \rfr{spherical} it turns out
 \eqi\sin i \left(\dert{i}{t}\right) =\left(\sin\Psi\cos I-\cos\Psi\sin I\cos\Omega\right)\dert{\Psi}{t}+\sin\Psi\sin I\sin\Omega \left(\dert{\Omega}{t}\right).\lb{eho}\eqf
 As \rfr{Gauss} shows for $\lambda\rightarrow\Psi$, only those  perturbing accelerations, like the general relativistic GE one, having no out-of-plane components $A_N$ leave $i$ unaffected since both $\Psi$ and $\Omega$ may change due to $A_N$. Thus, in the following expressions for the effects due to the general relativistic GM field and $Q_2$ it must be recalled that $\Omega$ is intended to be referred to the equatorial plane, not to the plane of the sky. It must also be noted that both $I$ and $\Psi$ are unknown.
\subsection{The effect of general relativity}
General relativity dynamically affects the radial velocity of non-circular orbits by causing averaged long-term variations of it. Indeed, an exact calculation in $e$ with \rfr{ge} yields
\eqi\left\langle\dot v_{\rho}^{\rm (GE)}\right\rangle = \left(n^2 {\mathcal{R}_g}\right)\rp{ 15 e(1+e^2)\sin i \sin\omega}{8\left(1-e^2\right)^{5/2}},\lb{velge}\eqf for the Schwarzschild-type, gravitoelectric component, while the secular acceleration due to  the Lense-Thirring-type, gravitomagnetic terms of \rfr{gm} is\footnote{
\ros{It is assumed that the orbital motion is prograde with respect to the BH's spin; a $-$ sign would occur in the opposite case.}
}
\eqi\left\langle\dot v_{\rho}^{\rm (GM)}\right\rangle = \left(\rp{nGL}{c^2 a^2}\right)\rp{e}{4 (1-e^2)^{2}}\left[\mathcal{V}_c(i,I,\Omega,\Psi)\cos\omega+\mathcal{V}_s(i,I,\Omega,\Psi)\sin\omega\right],\lb{velgm}\eqf
with
\begin{equation}
{\begin{array}{lll}
\mathcal{V}_c &\doteq & 11\cot i \sin I\sin\Psi\sin\Omega,\\ \\
\mathcal{V}_s &\doteq &\rp{\csc i}{4}\left\{\cos\Omega\sin 2I\left(\sin\Psi -\sin 3\Psi\right)
-\right.\\ \\
&-& \left. \sin^2\Psi\cos\Psi\left[\cos 2I \left(3 + \cos 2\Omega\right) +
        2\sin^2\Omega\right]+ 104\sin^2 i \cos\Psi
\right\}.
\end{array}}\lb{coeffvelgm}
 \end{equation}
 Let us recall again that  $I$ is the angle-unknown-between $\bds L$ and the line-of-sight, and $\Omega$ lies in  the equatorial plane of the source: for it the values determined in literature \citep{Ghez,Gille,Gille2} cannot be used because they refer to the plane of the sky\footnote{In principle, spherical trigonometric formulas relate both the nodes $\Omega_{\rm e}$ and $\Omega_{\rm s}$, where e and s denotes the equator and the sky. However, the knowledge of the angle, in the plane of the sky, between the equatorial plane and the reference $\{xz\}$ plane would be required.}. Also $\Psi$ is unknown. Anyway, in order to give an order-of-magnitude evaluation of the Lense-Thirring long-term radial acceleration we will simply look at the multiplicative dimensional factor in front of the square brackets in \rfr{velgm}.
For $e\rightarrow 0$ both \rfr{velge} and \rfr{velgm} vanish.
The general relativistic effects are non-vanishing either for edge-on ($i=90$ deg) or equatorial orbits ($\Psi=0$), with \rfr{velge} which is independent of $\Psi$, contrary to \rfr{velgm}.

Some preliminary and approximate calculation of the effects of both the general relativistic GEM dynamical effects on the velocity of S stars orbiting the SBH in GC can be found in \citet{Khan}; they do neither  deal with the directly measurable radial velocity nor with its possible variations. See also \citet{Ange} for GEM effects on the traveling photons paths.

Special relativistic effects on the radial velocity related to the Doppler effect have been considered by \citet{Zuck}; anyway, they do not involve net variations of $v_{\rho}$ over one orbital revolution.
\subsection{The quadrupole mass moment}
The BH's oblateness causes an averaged long-term variation of the radial velocity only if the orbit is elliptic. Indeed, it turns out
\eqi \left\langle\dot V_{\rho}^{ (J_2)}\right\rangle = -\left(\rp{Q_2^{\bullet}}{a^4}\right)\rp{3e  }{32 \left(1-e^2\right)^{7/2} \sin i}\left[\mathcal{J}_c(e,i,I,\Omega,\Psi)\cos\omega+\mathcal{J}_s(e,i,I,\omega,\Omega,\Psi)\sin\omega\right],\lb{velj2}\eqf
with
\begin{equation}
{\begin{array}{lll}
\mathcal{J}_c &\doteq & 10(1-e^2)\cos i \sin I\sin 2 \Psi\sin\Omega,\\ \\
\mathcal{J}_s &\doteq & 2\left(1-e^2\right)\cos i\sin 2\Psi\left(\cos I\sin\Psi-\sin I\cos\Psi\cos\Omega\right)+\\ \\
&+& \sin^2 i \left[7+47\cos 2\Psi+\sin^2\Psi\cos 2\omega-\right.\\ \\
&-&\left.\rp{e^2}{16} \left(259+429\cos 2\Psi-44\sin^2\Psi\cos 2\omega\right)\right].
\end{array}}\lb{coeffvelj2}
 \end{equation}
 It is an exact result in $e$, and vanishes in the limit $e\rightarrow 0$. Note that, according to \rfr{coeffvelj2}, \rfr{velj2} vanishes neither for equatorial orbits ($\Psi=0$) nor for edge-on configurations ($i=90$ deg). Also in this case, $\Omega$ refers to the equatorial plane.
 \subsection{The diffuse inner dark matter}
 The averaged long-term effect of the diffuse inner dark matter\footnote{The effects of non-baryonic dark matter on the perinigricon precession have been considered by, e.g., \citet{Zak}.} on $\dot v_{\rho}$ can be worked from \rfr{dmacc}. An approximate calculation with
 \eqi \left(1+\rp{r^2}{d^2_{\rm c}}\right)^{-3/2}\approx 1-\rp{3}{2}\rp{r^2}{d^2_{\rm c}}\lb{approz}\eqf and
 \eqi (1+e\cos f)^{-4}\approx 1-4 e\cos f,\ (1+e\cos f)^{-5}\approx 1-5 e\cos f,\eqf
 yields
 \eqi
 \left\langle\dot v_{\rho}^{\rm (dm)}\right\rangle = \left(\rp{GM_{\bullet} a}{d^3_{\rm c}}\right)\rp{e(1-e^2)^{3/2}\mu\sin i}{8}\left[
 32-5e^2(2+3e^2)+ 6(1-e^2)^2(5e^2 -11)\rp{a^2}{\ros{d}^2_{\rm c}}\right]\sin\omega.\lb{dmradvel}
 \eqf
Note that \rfr{dmradvel} vanishes for $e\rightarrow 0$ and for face-on ($i=0$) orbital configurations. Concerning the validity of the approximation of \rfr{approz}, it actually holds for S2. Indeed, its orbital parameters and \rfr{densita} for $d_{\rm c}$ tell us that
\eqi 0.002\leq \rp{r^2}{d^2_{\rm c}} \leq 0.4.\eqf
\section{Numerical evaluations and confrontation with the measurement accuracy}\lb{numerizi}
By using the known orbital parameters of S2 along with the associated uncertainties \citep{Gille2}, the known mass $M_{\bullet}$ of the SBH in the GC \citep{Gille2} and the  values for its angular momentum and quadrupole mass moment from \rfr{spinbh} and \rfr{sbozo}, it turns out that\footnote{The figures for $\left\langle\dot v_{\rho}^{\rm (GM)}\right\rangle$ and $\left\langle\dot v_{\rho}^{ (Q_2)}\right\rangle$ refer to the dimensional multiplicative factors in front of the square brackets in \rfr{velgm} and in \rfr{velj2}.}
\eqi
\begin{array}{lll}
\left\langle\dot v_{\rho}^{\rm (GE)}\right\rangle & = & (8.2\pm 2.6)\times 10^{-5}\ {\rm m\ s}^{-2},\\ \\
\left\langle\dot v_{\rho}^{\rm (dm)}\right\rangle & = & 3.8\times 10^{-6}\ {\rm m\ s}^{-2},\\ \\
\left\langle\dot v_{\rho}^{\rm (GM)}\right\rangle &\propto & 1.3\times 10^{-8}\ {\rm m\ s}^{-2}, \\ \\
\left\langle\dot v_{\rho}^{(Q_2)}\right\rangle &\propto & 1\times 10^{-10}\ {\rm m\ s}^{-2}.
\end{array}\lb{numeri}
\eqf
Concerning a possible measurement of a net change in the radial velocity of S2 after it completed  one full orbital revolution, no empirical evidences for it exist to date in literature, at least to the knowledge of this author. Anyway, measurements of the magnitude of the three-dimensional acceleration of S2 after 2 years (1997-1999) are available; its accuracy amounts to $4\times 10^{-4}$ m s$^{-2}$ \citep{Ghez00}. By assuming an uncertainty of about 15 km s$^{-1}$ in measuring the radial velocity of S2 \citep{Gille}, an overall accuracy of the order of $2.4\times 10^{-5}$ m s$^{-2}$ in $\left\langle \dot v_{\rho}\right\rangle$ may  be assumed in future over an observational time span $\Delta t=20$ yr. Actually, it must be pointed out that the currently available radial velocity measurements do not yet cover one full orbit revolution for S2. Indeed, the first radial velocity
data points are from 2000, then 2002; they are more densely sampled from
2003 onwards (S. Gillessen, private communication, August 2010).

It has to be pointed out that  the total accuracy reachable in the changes in the radial velocities is actually impacted by the uncertainty in LSR itself. Indeed, as explained by \citep{Ghez}, to obtain
the radial velocities with respect to the LSR, each observed radial velocity has to be corrected
for the Earth's rotation, its motion around the Sun, and
the Sun's peculiar motion with respect to the LSR (nominal value $U_{\odot}
= 10$ km s$^{-1}$, directed radially inwards, \citep{Dehe}).
Since the LSR is defined as the velocity of an object in
circular orbit at the radius of the Sun, the Sun's peculiar
motion with respect to the average velocity of stars in
its vicinity should give the Sun's motion toward the GC.
In all such a machinery, also the rotation speed $\Theta_0$ of LSR enters; recent evaluations by \citet{Reid} yield an uncertainty of the order of 16 km s$^{-1}$ corresponding to a future uncertainty of $2\times 10^{-5}$ m s$^{-2}$ over $\Delta t=20$ yr.
Moreover, also  the motion of the SBH itself should be taken into account \citep{Ghez}. In particular, the uncertainty in its radial velocity  can be evaluated to be 2 km s$^{-1}$ \citep{Gould} implying a limit in the accuracy in $\left\langle\dot v_\rho\right\rangle$ of about $1\times 10^{-6}$ m s$^{-2}$ over $\Delta t=20$ yr. Such limiting factors should be taken into account when future improvements in measuring radial velocities are discussed.
Searches for pulsars orbiting the SBH are currently underway \citep{puls}; their discovery may yield orbiting probes with  a better accuracy in their radial velocity changes.

These considerations show that while the gravitomagnetic and the quadrupole effects are far from being directly detectable in such a way, the gravitoelectric trend lies just at the edge of the measurability capabilities. The effect of the diffuse dark matter inside the orbit of S2 is one order of magnitude smaller than the general relativistic GE one. Incidentally, such figures indicate that higher-order corrections to the computed effects of \rfr{velge}-\rfr{velgm} in the linear GEM approximation  due to the relativistic motion of S2 can be neglected at the moment.

\ros{
Finally, in view of the possible discovery in the near future of stars closer than S2 we plot in Figure \ref{figura} the magnitudes of the effects considered
as functions of the semi-major axis for different values of the eccentricity.
\begin{figure*}
\centering
\begin{tabular}{cc}
\epsfig{file=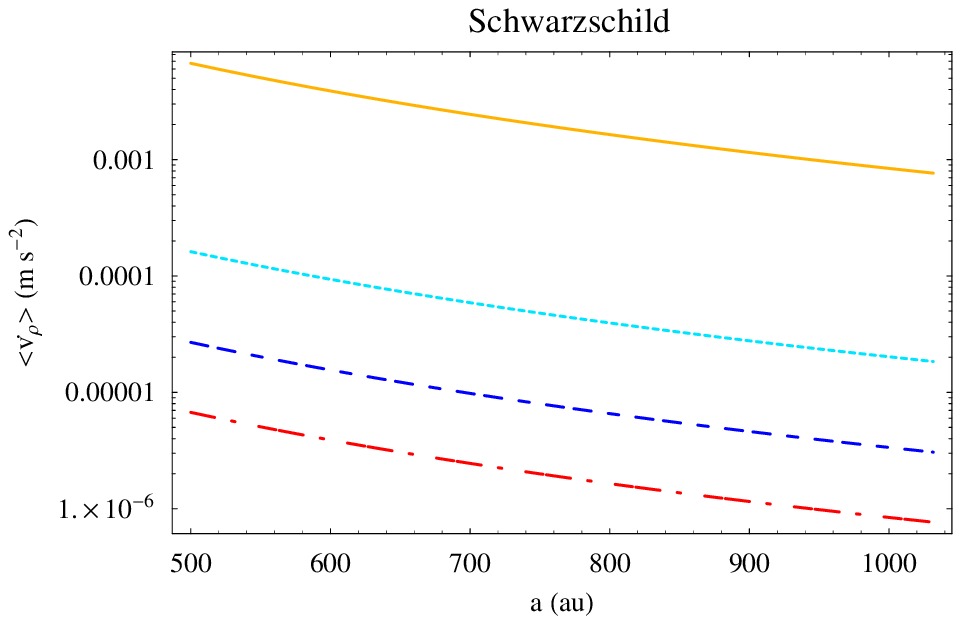,width=0.45\linewidth,clip=} &
\epsfig{file=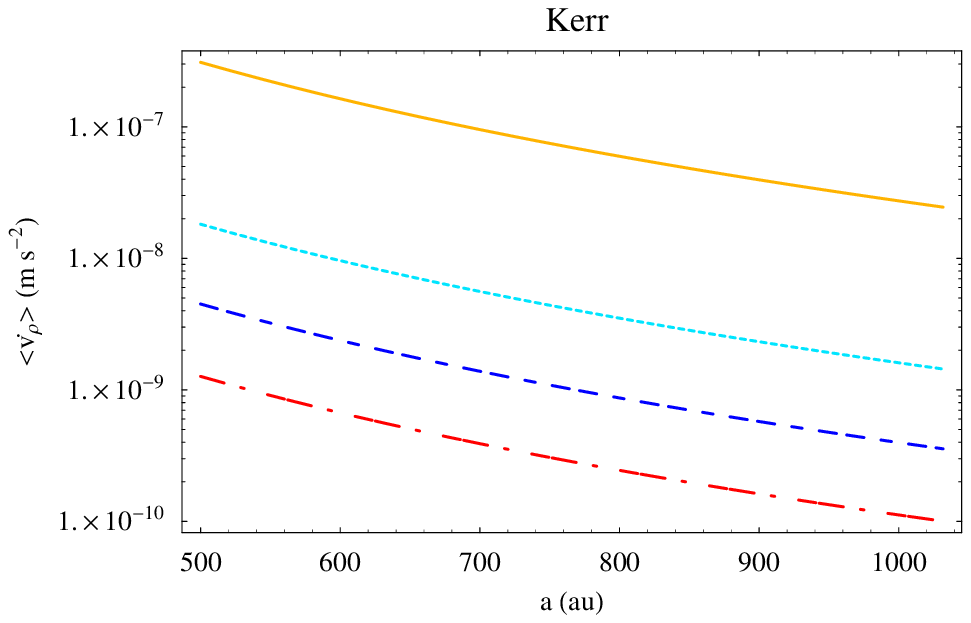,width=0.45\linewidth,clip=} \\
\epsfig{file=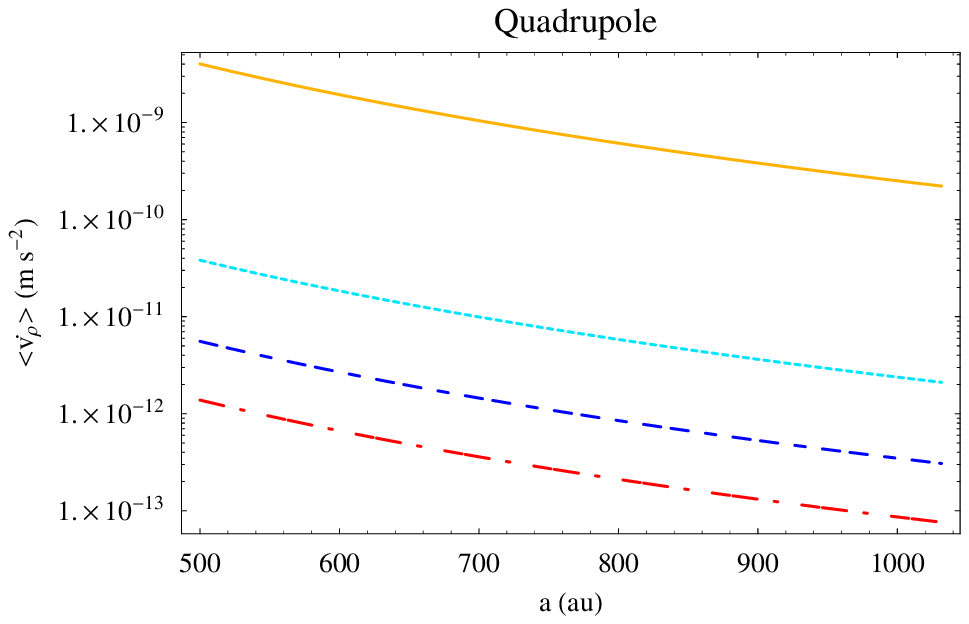,width=0.45\linewidth,clip=} &
\epsfig{file=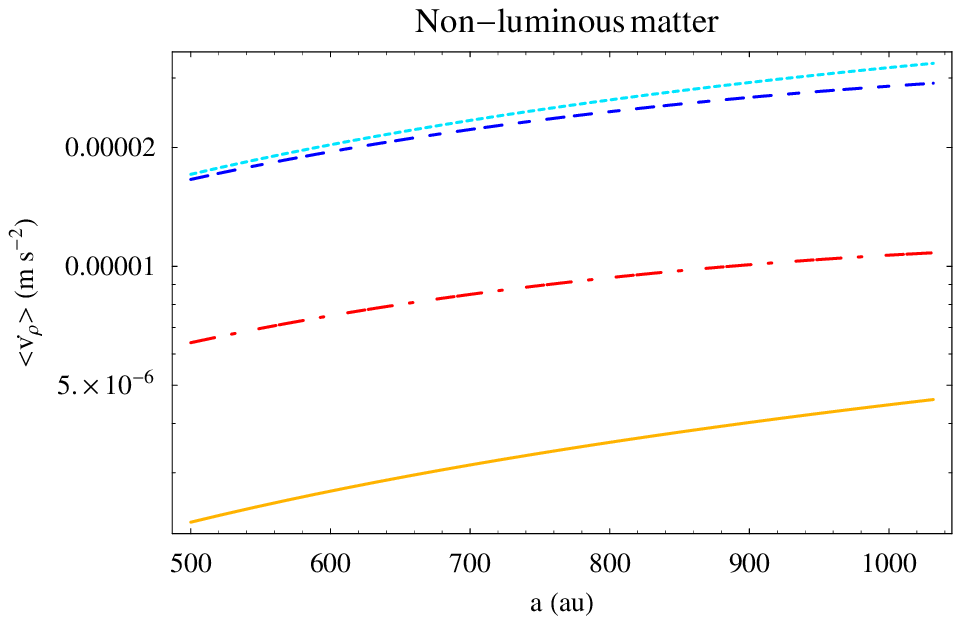,width=0.45\linewidth,clip=}
\end{tabular}
\caption{Maximum values of the long-term time variations $\left\langle\dot v_{\rho}\right\rangle$, in m s$^{-2}$, as a function of $a$ ($500\ {\rm au}\leq a\leq 1031.69\ {\rm au}$) for different values of the eccentricity: $e=0.1$ (red dash-dotted line), $e=0.3$ (blue dashed line), $e=0.6$ (light blue dotted line), $e=0.9$ (yellow continuous line). For the  the spin $L_{\bullet}$ and the quadrupole $Q_2^{\bullet}$ of the SBH we used the values of \rfr{spinbh} and \rfr{sbozo}, respectively. }\lb{figura}
\end{figure*}
}
\section{Conclusions}\lb{conclu}
One of the directly measurable quantities of the system constituted by the S stars orbiting the Supermassive Black Hole  located at the center of the Milky Way in the radio source \sgr is the radial velocity. Given that S2 has already completed one full orbital revolution, with a period of 15.9 yr, since its discovery, and in view of the possible detection in the near future of other stars and pulsars with shorter orbital periods, we looked at the cumulative, long term time variations of the radial velocity caused by several Newtonian and Einsteinian dynamical effects. They are both the general relativistic Schwarzschild and Kerr-like components of the spacetime metric, the quadrupole mass moment and the diffuse dark mass distribution made by stellar remnants enclosed within the star's orbit. We analytically worked out the long-term variations in the radial velocity induced by them by finding non-zero effects for all of them. We used S2 for numerically computing their magnitudes. They are $8\times 10^{-5}$ m s$^{-2}$ (Schwarzschild), $4\times 10^{-6}$ m s$^{-2}$ (dark matter), $1\times 10^{-8}$ m s$^{-2}$ (Kerr), $1\times 10^{-10}$ m s$^{-2}$ (quadrupole), respectively. In computing the general relativistic variations of the radial velocity, we remained within the post-Newtonian regime by neglecting relativistic corrections of higher order in the equations of motion. The figures for the the Kerr and quadrupole effects have been computed by using the latest determinations of the angular momentum parameter of the Galactic black hole, and in the \virg{no-hair} hypothesis, respectively. For the dark matter distribution we used a Plummer-like mass density profile. By assuming a present-day uncertainty of about 15 km s$^{-1}$ in the radial velocity measurements, its time changes may be detected in the future at a $\approx 10^{-5}$ m s$^{-2}$ level over an observational time span of 20 yr; at present, radial velocity data cover just 7 yr. Even if such evaluations will turn out to be not too optimistic, a detection of the Kerr and the quadrupole-induced cumulative changes of the radial velocity  seems to be unfeasible.

\section*{Acknowledgments}
I thank S. Gillessen for useful information about the radial velocity and QPOs. I am also grateful to an anonymous referee for her/his useful comments which
improved the manuscript.

\end{document}